\documentclass[acmsmall]{acmart}

\AtBeginDocument{}

\setcopyright{acmcopyright}
\copyrightyear{2018}
\acmYear{2018}
\acmDOI{XXXXXXX.XXXXXXX}

\acmVolume{37}
\acmNumber{4}
\acmArticle{111}
\acmMonth{8}

\usepackage[normalem]{ulem} \usepackage{graphicx}
\usepackage[breakable,skins]{tcolorbox}

\usepackage{todonotes}
\usepackage{url}

\clearpage{}\usepackage[normalem]{ulem} \usepackage{xcolor}

\usepackage{ifthen}
\newboolean{showcomments}
\setboolean{showcomments}{true} \ifthenelse{\boolean{showcomments}}
  {\newcommand{\nb}[2]{
    \fcolorbox{gray}{yellow}{\bfseries\sffamily\scriptsize#1}
    {\sf\small$\blacktriangleright$\textit{#2}$\blacktriangleleft$}
   }
   
  }
  {\newcommand{\nb}[2]{}
   
  }

\clearpage{}

\begin{document}
\title{{Aspects of} Modelling {Requirements} {in Very-Large Agile Systems Engineering} }

\author{Grischa Liebel}
\email{grischal@ru.is}
\orcid{1234-5678-9012}
\affiliation{\institution{School of Technology, Reykjavik University}
  \streetaddress{Menntavegur 1}
  \city{Reykjavik}
  \country{Iceland}
  \postcode{102}
}

\author{Eric Knauss}
\email{eric.knauss@cse.gu.se}
\orcid{0000-0002-6631-872X}
\affiliation{\institution{Department of Computer Science and Engineering, Chalmers $|$ University of Gothenburg}
  \city{Gothenburg}
  \country{Sweden}}

\begin{tcolorbox}
\textbf{Self-archiving note:}\\
This is a pre-peer-review version of an article currently under submission to a journal in Software Engineering.
\end{tcolorbox}

\begin{abstract}
Using models for requirements engineering (RE) is uncommon in systems engineering, despite the widespread use of model-based engineering in general.
One reason for this lack of use is that formal models do not match well the trend to move towards agile developing methods.
While there exists work that investigates challenges in the adoption of requirements modeling and agile methods in systems engineering, there is a lack of work studying successful approaches of using requirements modelling in agile systems engineering.
To address this gap, we conducted a case study investigating the application of requirements models at Ericsson AB, a Swedish telecommunications company.
We studied a department using requirements models to bridge agile development and plan-driven development aspects.
We find that models are used to understand how requirements relate to each other, and to keep track with the product's evolution.
To cope with the effort to maintain models over time, study participants suggest to rely on text-based notations that bring the models closer to developers and allow integration into existing software development workflows.
This results in tool trade-offs, e.g., losing the possibility to control diagram layout.
\end{abstract}



\maketitle

\section{Introduction}
\label{sec:intro}
Driven by success stories in small-scale software development, agile development is increasingly adopted in large-scale software and systems engineering \cite{dikert16,lagerberg13,salo08,eklund14,berger15}.
However, context factors such as long lead times \cite{berger15}, safety criticality \cite{kasauli18}, and the scale of development itself make this adoption challenging.
In particular, challenges relate to Requirements Engineering (RE), such as building and maintaining a shared understanding of customer value and the system requirements \cite{kasauli17,kasauli21}.

To build and maintain system knowledge over time, models have been used as a suitable means of documentation \cite{kasauli21}.
Specifically, models are often cited as a way to deal with complexity that arises from the scale of systems \cite{selic1998using}.
However, while the use of models is common in systems engineering \cite{liebel18survey}, using requirements models is uncommon in practice \cite{liebel18sosym,loniewski10}.
In the context of large-scale agile software and systems engineering, we are not aware of any work investigating the use of models in industry.

Therefore, the goal of this paper is to better understand the potential of using requirements models in very large-scale (VLS) agile \cite{dingsoyr2014large} systems engineering.
To do so, we conducted a case study of a single department at Ericsson AB, a large Swedish telecommunications provider, which has long-ranging experience using requirements models in a VLS agile setting.
We aim to answer the following research questions (RQs).
\begin{itemize}
\item[\textbf{RQ1:}] What sentiments exist for and against the use of requirements models in VLS agile systems engineering?
\item[\textbf{RQ2:}] How do different stakeholders use requirements models in VLS agile systems engineering?
\item[\textbf{RQ3:}] What are the needs to support the intended use of requirements models in VLS agile systems engineering?
\end{itemize}

To answer these questions, we collected survey data, followed up with a number of semi-structured interviews to find answers to patterns observed in the survey.

We find that the requirements models at the case department serve as a \emph{boundary object} that relates the agile world in individual teams with the overall waterfall-like process that deals with product requirements and their long-term evolution.
While engineers are positive regarding the use of models, many take a practical stance concerning the feasibility of continuously maintaining these models over time.
To achieve an updated and maintained model, text-based modelling approaches such as PlantUML\footnote{\url{https://plantuml.com/}} with certain inherent limitations such as automatic layouting are seen as inevitable. 
Furthermore, to avoid deterioration of models over time, our study participants suggest generating simple artefacts from the models, e.g., documentation.
This would encourage engineers to regularly update the models, as derived artefacts would otherwise become outdated. \section{Related Work}
\label{sec:related_work}

There exists a broad body of work on the use of models in industry, and suggestions on how to use models for RE-related activities.
In the following, we will discuss this work in detail.

\subsection{Use of Models in Industry}
There are numerous case studies reporting how models are used in industry, e.g., \cite{baker05a,mohagheghi13a,hutchinson11a,hutchinson11b,whittle13,liebel18sosym}.

In a case study at Motorola, Baker et al.~\cite{baker05a} discuss how Model-Based Engineering (MBE) is used at Motorola over a period of 20 years.
The authors report several positive effects, such as defect reductions and increases in productivity, but also a lack of tools and tool interoperability, poor performance of generated code, and a lack of scalability of the modelling approach.

Experiences from three European companies with MBE techniques and tools are presented by Mohagheghi et al.~\cite{mohagheghi13a} in terms of a qualitative study.
The authors find that simulation and testing opportunities are positive aspects of using MBE, while tool problems and the complexity of models are listed as drawbacks.

Hutchinson et al. study the use and adoption of MBE in industry in a series of qualitative and mixed-methods studies \cite{hutchinson11a,hutchinson11b,whittle13,hutchinson14}.
The overall finding of this study series is that the organisation context and several non-technical topics need to be considered for MBE to succeed.
For instance, the authors report that significant additional training is needed for the use of MBE.
From their interviews, the authors conclude that especially people's ability to think abstractly seems to have significant impact on their ability to create models.
In addition, several technical challenges such as tool shortcomings impede the use and success of MBE.

In a case study at two automotive companies~\cite{liebel18sosym}, we find that models are used in automotive RE to improve communication and to handle complexity.
However, stakeholders prefer informal models and whiteboard sketches over formal modelling notations.

\subsection{Frameworks for Using Models During RE}
Several frameworks and methods have been suggested that include the use of models for or during RE.
 
Pohl et al.~\cite{pohl12} introduce the SPES 2020 Methodology for the development of embedded systems.
During RE in particular, the framework suggests a separation between \emph{solution-independent} and \emph{solution-oriented} diagrams. 
Practical experiences with SPES are reported in \cite{bohm14b} and \cite{brings17}.
In \cite{bohm14b}, B\"{o}hm et al. present their experiences with SPES in an industrial project at Siemens.
The authors apply SPES to a mature, already running train control system, using a specification of ``high quality''.
Findings are that ``the high quality of input documents, and cooperation with product experts were considered the most influential success factors''.
Brings et al.~\cite{brings17} discuss experiences of using SPES in the area of cyber-physical systems.
The authors report that they ``identified problems resulting from an increased number of dependencies.'' and ``the need to cope with redundancies caused by properties which are system as well as context properties in a structured manner.''.

Apart from the SPES framework, there exist several proposed processes and frameworks for requirements modelling, e.g., \cite{vogelsang14,brandstetter15,braun14,fockel14,berenbach12}.

Vogelsang et al.~\cite{vogelsang14} propose to model requirements and architecture in parallel, and evaluate the approach with 15 master students.
In particular, the authors propose the use of Message Sequence Charts.

Brandstetter et al.~\cite{brandstetter15} present a process to perform early validation of requirements by means of simulation, using the control software of a desalination plant as an industrial case.
Experiences of using the approach are discussed, but details on the execution of the use case are largely missing.

Resulting from a research project with academic and industrial partners, Braun et al.~\cite{braun14} propose the use of model-based documentation.
For RE, these include goal models, scenario models and function models. 
To our knowledge, the approach has not been evaluated in terms of an empirical study.

Berenbach, Schneider, and Naughton~\cite{berenbach12} list several requirements they consider essential for a requirements modelling language, such as distinction between process and use case modelling.
The authors argue that using UML for requirements modelling has proven to be frustrating.
URML is piloted in one commercial project at Siemens, showing that the proposed concepts are useful.

Finally, the Model-Driven Requirements Engineering (MoDRE) workshop series that has taken place since 2011 contains many contributions on how models, in particular in the context of model-driven development, can be used for RE purposes.

\subsection{RE in Large-Scale Systems Engineering}

Initially, agile approaches were focused on small teams developing software \cite{Beck1999,Meyer2014,Kahkonen2004}. 
The success of these approaches have led to their adoption at scale \cite{Dikert2016,lagerberg13,Salo2008}, where {non-agile, plan-driven, and stage-gate based processes} have been the norm \cite{Pernstal12}.

Due to their iterative nature, agile approaches are suitable for building systems whose requirements may change; further, experience from early versions of a system can impact later versions \cite{Beck1999,Meyer2014,Gren2020}. 
Gren and Lenberg even argue that the main motivation for choosing agile methods is to be able to respond to changing requirements \cite{Gren2020}.

However, Heikkil\"{a} et al.~\cite{heikkila2015mapping} find in their mapping study that there is no universal definition of agile RE. 
Instead, they report that requirements-related agile practices such as the use of customer representatives, prioritization of requirements, or growing technical debt are particularly hard to manage.
The same authors also present a case study at Ericsson, where they investigate the flow of requirements in large-scale agile~\cite{heikkila2017managing}.
They find that practitioner perceive benefits such as increased flexibility, increased planning efficiency, and improved communication effectiveness.
However, the authors also report problems such as overcommitment, organizing system-level work, and growing technical debt.
In their case study on the use of  agile RE at scale, Bjarnason et al.~\cite{bjarnason2011case} also report that agility can mitigate communication gaps, but at the same time may cause new challenges, such as ensuring sufficient competence in cross-functional teams.
In a case study with 16 US-based companies, Ramesh{ et al.}~\cite{ISJ:ISJ259} identify risks with the use of agile RE such as neglecting non-functional requirements or customer inability.
A systematic literature review on agile RE practices and challenges reports eight challenges posed by the use of agile RE~\cite{inayat2015systematic}, such as customer availability or minimal documentation.
However, the authors also report 17 challenges from traditional RE that are overcome by the use of agile RE.
The authors conclude that there is more empirical research needed on the topic of agile RE.
Consequently, Kasauli et al. \cite{kasauli21} report on RE challenges in scaled-agile system development that are neither addressed in contemporary RE literature nor by established frameworks for scaled-agile.

Paetsch \cite{paetsch2003requirements} suggest that agile methods and RE are pursuing similar goals in key areas like stakeholder involvement and therefore could be integrated in a good way.
The major difference is the emphasis on the amount of documentation needed in an effective project.
{Meyer, in contrast, criticizes agile methods for limiting requirements engineering to functional requirements described through (exemplary) scenarios and discouraging upfront planning \cite{Meyer2014}. 
In fact, in practice such functional requirements are often described as user stories, e.g. formulated as boilerplate statements: ``As a $<$role$>$ I want $<$feature$>$ so that $<$value$>$.'' \cite{Leffingwell2010}. 
The much more detailed requirements of plan-driven approaches are omitted; instead, agile methods push for a continuous dialogue (with customer representatives or product owners) and comprehensive sets of tests, which are ideally automated \cite{Meyer2014}.

Given the set of challenges with managing requirements in scaled agile, it is unlikely that user stories and automated tests are enough to enable a shared understanding of requirements in agile at scale. 
It is therefore that we investigate the use of requirements models in agile.

\subsection{Models in Agile Development}
As a final area of related work, several authors have explored how models can be used in agile development, e.g., \cite{rumpe2002agile,ambler2001agile,hansson2014}.

Ambler~\cite{ambler2001agile} argues that modelling and agile development can go hand in hand.
The author describes important aspects to succeed with agile modelling, e.g., using as simple tools as possible, fostering effective communication, and building agile modelling teams.
Similarly, Rumpe~\cite{rumpe2002agile} argues that modelling can be used as a part of agile methodologies to further increase development efficiency.
Concretely, the author suggests to use models for code and test case generation.

A number of further approaches to use models during agile development have been proposed as a part of the Extreme Modeling (XM) workshop series.
However, as noted by Hansson et al.~\cite{hansson2014}, existing work on agile modelling suffers from a lack of empirical evidence on its application in industry.

In summary, there is a large body of work on how models are used in industry, including benefits and challenges of using models.
Additionally, challenges of agile development and agile RE at scale are studied in considerable depth.
Finally, a substantial amount of solution proposals for using models for RE and during agile development exist.
However, to our knowledge there are no detailed studies investigating industry cases of successful model use for RE activities. \section{Research Method}
\label{sec:method}
To address the RQs, we conducted a case study at a department in a large Swedish telecommunications company - in the following referred to as the case department/company.
We embrace a constructivist world view, emphasising that different engineers at the case department have subjective views and opinions on the topic under investigation.
The case study is both exploratory and confirmatory in nature.
That is, we use a set of propositions we formulated initially and updated throughout the study.
At the same time, we included a number of open questions to be investigated as part of the study.

\subsection{Case Description}
We conducted this study in one department at Ericsson AB, a large Swedish telecommunications provider.
In that department, more than 30 Scrum teams develop a single product in parallel based on a scaled agile approach.
Cross-functional teams independently work on backlog features all the way to delivery on the main branch.
Specialised coordination roles exist, e.g., for integration or architecture tasks.
Scrum sprints are based on a backlog and a hierarchy of product owners breaks down product requirements and customer visible features to backlog items.
While these product owners represent the customer requirements towards the product development, \emph{system managers (SMs)} represent a system requirements perspective.
These SMs also interact with agile teams in providing the system-level knowledge.
Further products are developed using a similar methodology. 

Hardware development at the company is largely decoupled from software development. New hardware becomes available with a regular, but low frequency.

The studied case department is a department at Ericsson AB.
At the time of the study, there were approximately 200 engineers working at the department.
Development is closely aligned with existing standards that describe technical solutions in much detail, e.g., \cite{3gpp_tr1601}.

Requirements on system level are stored in the tool T-Reqs~\cite{knauss2018t}, which has been developed in house.
T-Reqs allows storing text-based requirements and other artefacts together with code in version control systems such as git, thus bringing these artefacts closer to developers~\cite{knauss2018t}.
The tool has been used at the case company since 2017.

Models are used to keep track of the system requirements and their relation.
This is primarily done in the form of UML activity diagrams, where activities denote requirements and the flow between activities their relation and order.
Models are created and maintained manually.
More details on the used models are presented in Section~\ref{sec:survey} and Section~\ref{sec:interviewProps}. 

\subsection{Study Scope and Propositions}
From previous research and initial scoping meetings with two contact persons at the case department, we formulated a number of propositions addressing the three research questions.
These are depicted in Table~\ref{tb:props}.
For each, we describe the origin of the proposition.

\begin{table}[!ht]
\centering
    \begin{tabular}{|p{.05\linewidth}|p{.6\textwidth}|p{.12\textwidth}|p{.1\textwidth}|}
    \hline
    Nr & Proposition & Source & RQs \\
            \hline
    P1 & Models are created by few experts, and mainly read by them. & \cite{liebel18sosym} & RQ2\\
    P2 & Access to models, and especially editing, is rare among testers and developers. & D & RQ2\\
P3 & Model creators are not modelling experts. Therefore, use of modelling languages is ad-hoc and varies across the organisation. & \cite{liebel18sosym} & RQ2\\
    P4 & Testers and developers do not see the need/use of modelling requirements. & D & RQ1, RQ2\\
     P5 & Only few diagram types (of the UML) are used. & \cite{liebel18sosym} & RQ2, RQ3\\
         P6 & Testers and developers do not think that the present models carry important information. & D & RQ1, RQ2\\
             P7 & Layouting of diagrams is important to the users. & DK, layout studies & RQ3\\
              P8 & Stakeholders believe that modelling should be integrated with existing development tools (e.g., git). & D & RQ1, RQ2\\
                  P9 & Stakeholders do not believe that the requirements models should be used for automated tasks. & \cite{liebel18sosym} & RQ1, RQ3\\
                      P10 & The current modelling solution is restricting employees in their work. & \cite{liebel18sosym} & RQ2\\
                          P11 & Even if a better/good modelling solution would be in place, most stakeholders would not update/maintain the model. & \cite{liebel18sosym} & RQ1, RQ3\\
                              P12 & Navigating between different diagrams is an important feature. & \cite{liebel18sosym} & RQ3\\
    \hline
    \end{tabular}
    \caption{Research Propositions, with sources and target RQs. D denotes \emph{Discussions}, DK denotes the authors' \emph{Domain Knowledge}.}
    \label{tb:props}
\end{table}
The propositions can be summarised as follows.
For RQ2, we envision an organisation in which few experts work with models (P1), while in particular the roles working with lower level of abstractions, testers and developers, do not use the models (P2), do not see the need for them (P4), and do not think that any relevant information is contained in the model (P6). People creating the models are not necessarily experts, resulting in an ad-hoc approach (P3). Few UML diagram types are in use (P5). The existing tool solution for modelling is restricting the employees in their work (P10).

For RQ1, we expect that sentiments towards modelling roughly resemble the current use: a few ``power users'' of models , but a substantial amount of people not believing in the usefulness of models.

For RQ3, we cover a few important tool decisions, including the need for only few diagram types (P5), the need for layouting capabilities (P7), automation support (P9) and navigation between diagrams (P12).
Furthermore, we expected some insights from participants that would not use the models even with better tools (P11), since they might have additional input on what would be the preferred format.
For the remaining feature space, we chose an exploratory approach asking several free-text questions to get additional input.

\subsection{Survey Design, Execution and Analysis}
To evaluate the propositions, we designed an online survey.
Our contact persons reviewed the survey design.

After review, our contacts sampled 54 people at the case department, all of which they judged to have sufficient knowledge of the model to answer our questions.
We received 33 answers, i.e., a return rate of 61.11\%.
The participants worked in 16 different areas of the case department, covering various tasks and product aspects, both from functional and non-functional perspective.
However, SMs were over-represented among the participants (22 out of 33 participants had an SM role).
Finally, the majority of participants had substantial work experience in the case department (depicted in blue bars in Figure~\ref{fig:participantsExperience}) and modelling experience (depicted in yellow bars in Figure~\ref{fig:participantsExperience}).
\begin{figure}
    \centering
    \includegraphics[width=.7\linewidth]{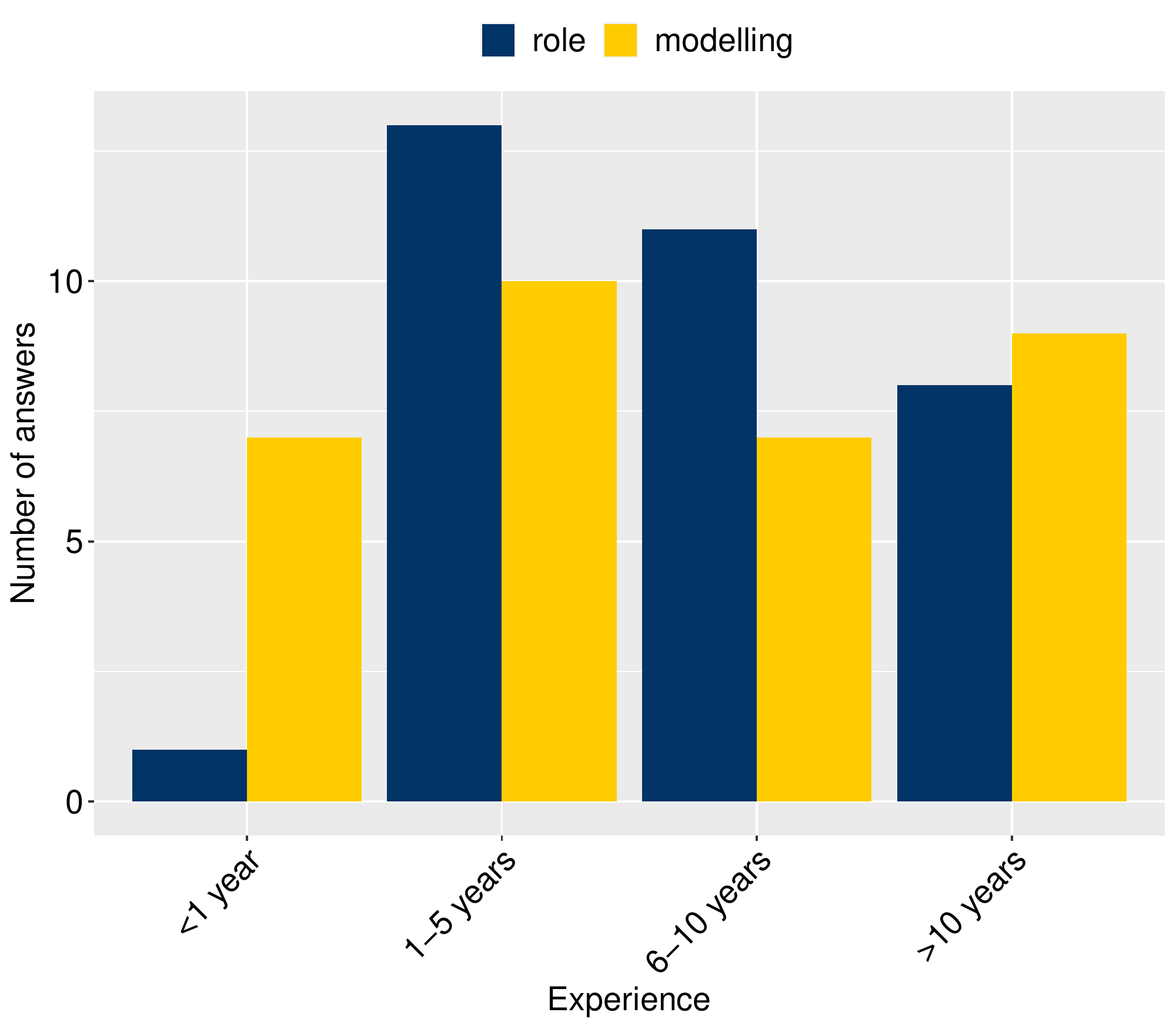}
    \caption{Model Creation and Reading Frequency}
    \label{fig:participantsExperience}
\end{figure}

We analysed the survey answers by creating summary statistics and evaluated the propositions in a qualitative manner, i.e., without employing statistical tests or related statistical methods.
The first author summarised open-ended questions by assigning topic codes~\cite{saldana15} to each stanza, then grouping related stanzas together and counting their frequency.
The mapping of survey questions to propositions is depicted in Table~\ref{tb:prop_to_q} in Appendix~\ref{app:prop_eval}.
As a form of member checking, we presented the results to our contact persons, who disseminated the findings in the department.

\subsection{Interview Follow-Up}
\label{sec:methodInterviews}
Following the questionnaire, we updated and refined our list of propositions and added some open questions (see Appendix~\ref{app:interviewProps}).
The open questions relate in particular to contradictions in the survey data.
For instance, while the majority endorsed using text-based models, the suggested solution does not support manual layouting, an important feature requested by the majority.
We used the proposition and questions as an input for the creation of the interview guide.

Our contact persons recruited five engineers to be interviewed.
We requested a varied set of roles and mindsets, to obtain diverse information.
In particular, we also asked them to recruit participants who might be skeptics of requirements modelling or modelling in general.
While this is a small sample, it nevertheless represents about 10\% of the survey sample size, i.e., engineers who are knowledgeable enough in modelling to answer our questions.

We analysed the interview transcripts using the following process.
Both authors, Grischa (GL) and Eric (EK), coded all interviews.
GL used a list of a-priori codes aligned with the propositions and questions, while EK used open coding.
In both cases, the coding followed a content coding approach \cite{saldana15}.
That is, we assigned codes that describe the content of the coded stanza, assigning codes on a per-answer basis.
In cases where the interviewees clearly discussed different content, we separated the answer into multiple stanzas which we coded differently.
GL piloted the initial a-priori codes on one interview, then modified them according to the pilot.
The final a-priori codebook is discussed in Appendix~\ref{app:codebook}.

After the first round of coding, we discussed the resulting code distribution and decided to continue a parallel approach.
That is, we jointly structured the existing codes in a second-cycle coding approach.
We hierarchically grouped the codes obtained from EK's open coding into the different (and much more abstract) a-priori codes, then grouped the resulting clusters according to our three research questions.
We then extracted candidate themes, which we validated using all stanzas coded with at least one of the open codes for the theme.
Simultaneously, GL analysed the interview data one more time and followed a holistic coding \cite{saldana15} approach, writing analytical memos while working through the data.
Finally, we integrated the initial themes from the second-cycle coding approach with the themes extracted from the holistic coding and memoing.

\subsection{Validity Threats}
Given the constructivist nature of this case study, we present the threats to validity in terms of transferability, credibility and confirmability \cite{petersen13}.

\subsubsection{Transferability}
Transferability describes to what extent results from the study can be transferred to cases that resemble the case under study \cite{petersen13}.

Many of the reported aspects are specific to the case department, e.g., the role of the \emph{SM} that connects agile teams with the system-level view.
However, we know from previous work \cite{kasauli21} that similar roles and situations exist in many systems engineering companies.
Therefore, we expect that the findings apply in similar cases as well.
One exception might be the large emphasis on software development at the case department, which is in contrast to many other systems engineering organisations, where hardware is developed in parallel and thus causes long lead times and longer feedback cycles.

We used purposeful sampling to select interviewees that had diverse background and at the same time could comment on the use of requirements models.
However, we did not reach saturation in all our themes.
This means that there might be additional facets or themes, or contrasting ideas that we did not capture.
This is a threat to the transferability of our findings.

\subsubsection{Credibility}
Credibility describes whether findings are reported truthfully, or have been distorted by the researchers \cite{petersen13}.

All interviews were recorded, and data analysis performed on the verbatim transcripts.
Additionally, we report quotes for all themes in our qualitative interview analysis.
This should ensure credibility of the findings.

We performed first-cycle coding and memo writing for both the free-text answers in the survey and the interview transcripts.
This should avoid threats to credibility arising from long chains of interpretation in our analysis.

\subsubsection{Confirmability}
Confirmability describes the extent to which conclusions made by researchers follow from the observed data \cite{petersen13}.

To structure our study, we used propositions prior to the survey and in between the survey and interviews.
We then evaluated them after each analysis step.
Furthermore, survey and interview instruments, as well as the codebooks are available in the appendix to this paper. \section{Exploratory Survey}
\label{sec:survey}

In the following, we present the results of the exploratory survey in terms of descriptive statistics and relations to the propositions.
We then discuss the implications of the survey findings.

\subsection{Survey Findings}
The resulting proposition evaluation is summarised in Table \ref{tb:prop_eval}.
\begin{table}[ht]
\centering
    \begin{tabular}{|p{.1\textwidth}|p{.65\textwidth}|p{.15\textwidth}|}
    \hline
    Number & Proposition & Supported by \newline Survey \\
        \hline
        P1 & Models are created by few experts, and mainly read by them.& Yes \\

        P2 &  Access to models, and especially editing, is rare among testers and developers. & Partially \\
P3 & Model creators are not modelling experts. Therefore, use of modelling languages is ad-hoc and varies across the organisation. & No\\
P4 & Testers and developers do not see the need/use of modelling requirements. & Partially\\
P5 & Only few diagram types (of the UML) are used. 
& Yes\\
P6 & Testers and developers do not think that the present models carry important information. & No\\
P7 & Layouting of diagrams is important to the users. & Yes\\
P8 & Stakeholders believe that modelling should be integrated with existing development tools (e.g., git). & Yes\\
P9 & Stakeholders do not believe that the requirements models should be used for automated tasks. They should instead be used as documentation only. & No\\
P10 & The current modelling solution is restricting employees in their work. & Partially\\
P11 & Even if a better/good modelling solution would be in place, most stakeholders would not update/maintain the model. & No\\
P12 &  Navigating between different diagrams is an important feature. & Yes\\
\hline
    \end{tabular}
    \caption{Evaluation of Propositions}
    \label{tb:prop_eval}
\end{table}

For P1, 5 participants state that they create/modify diagrams at least weekly (see the dark blue bars in Figure~\ref{fig:creationReadingFrequency}). Three of these participants are System Managers (SMs), one is a Developer and one is both a Developer and SM. When consulting the read access/use of diagrams (light yellow bars Figure~\ref{fig:creationReadingFrequency}), these five participants have weekly (4 answers) or daily (1 answer) read access to the diagrams. In the entire sample, only 2 more people stated that they read/use the diagrams on a weekly basis. This seems to confirm our proposition that it is indeed a small group responsible for modification and use of diagrams.

\begin{figure}
    \centering
    \includegraphics[width=.7\linewidth]{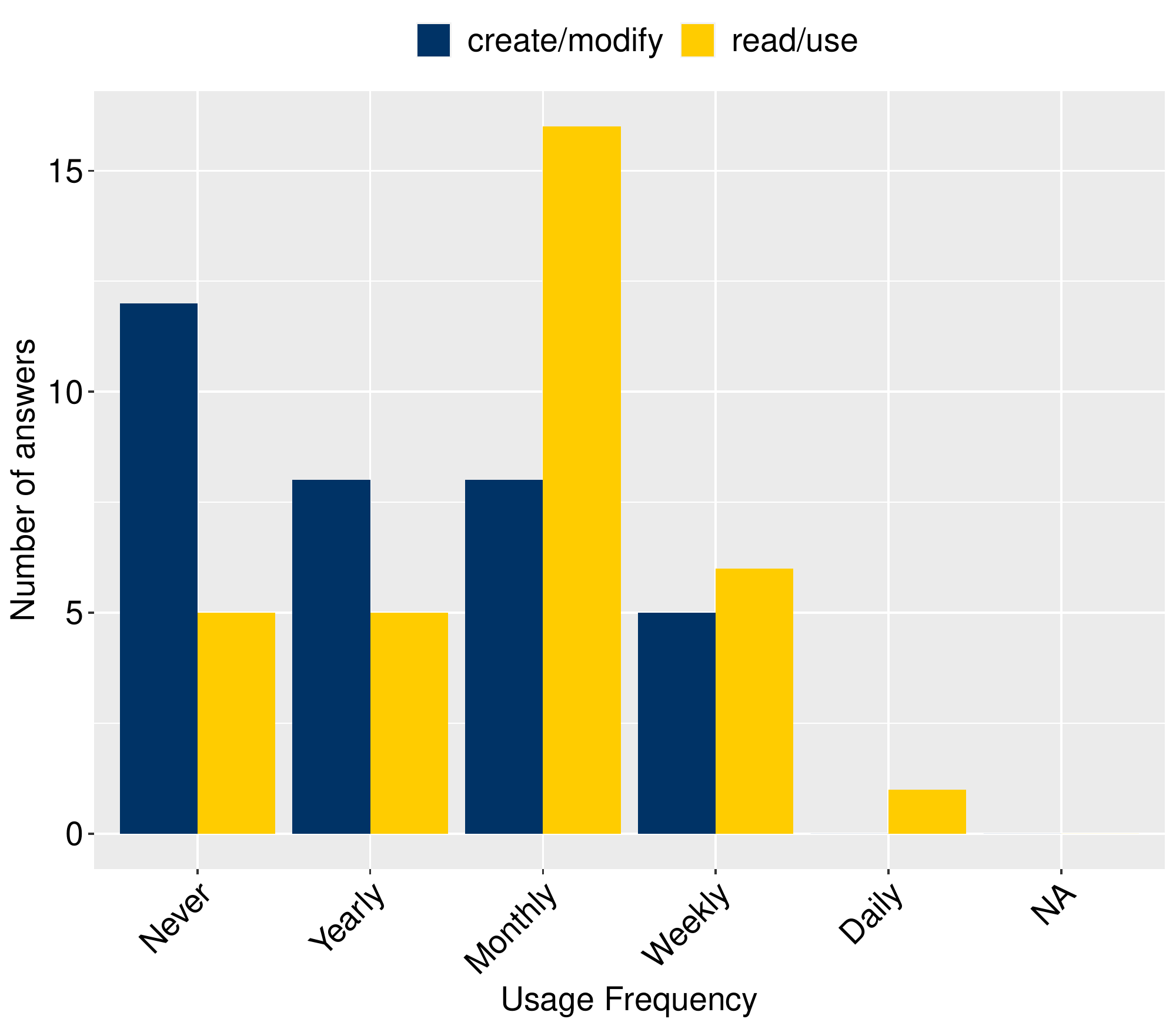}
    \caption{Model Creation and Reading Frequency}
    \label{fig:creationReadingFrequency}
\end{figure}

Our data shows a mixed picture for P2 (testers and developers access and modify models rarely).
Of the 8 people with development or testing roles (out of 33 participants, see Section~\ref{sec:method}), 5 state that they read models on a monthly basis, 2 weekly, and 1 yearly. Creation is less common, with 3 stating that they never create or modify models, 2 yearly, 1 monthly, and 2 weekly. This picture does not change significantly if we consider only those who have a pure testing/development role (without addition of Designer/Architect or System Manager). Overall, these figures do not allow a clear answer as to whether P2 is confirmed or not.
The free text answers indicate that people not accessing the models are mainly concerned with the tooling (difficulty of tooling, access to the tool) and the effort it takes to comprehend the models (too much detail, information spread across model layers/navigation).

P3 (model creators are not modeling experts) is not supported by our data and must be rejected. 
The participants who create/modify models at least weekly all have considerable experience with modelling (at least 5 years). 
We did however not ask whether they have a formal education, or proceed in an ad-hoc manner.

For P4 (testers and developers do not see the need of modeling requirements), the survey data shows again a mixed picture (parts of Figure~\ref{fig:agreementStatements}). 
Three survey participants agree or strongly agree that they would update the models regularly if they had a better tool. 
However, two participants strongly disagree and three do not know. 
There is again no noticeable difference between the pure developer/tester roles and others. 
Additionally, we do not see a pattern in the answers with respect to how the creation/modification patterns look like at the moment (e.g., ``Participants who already modify/create diagrams often would not do it more often''). 
Interestingly, six people are generally positive towards modelling, and the remaining two neutral. 
No one opposes modelling per se.
The free-text answers for this question do not give a clear justification of the pattern, either. 
However, one participant stated the concern that the current model is unreliable and therefore not useful, suggesting to assign someone to manage the model.

\begin{figure}
    \centering
    \includegraphics[width=1\linewidth]{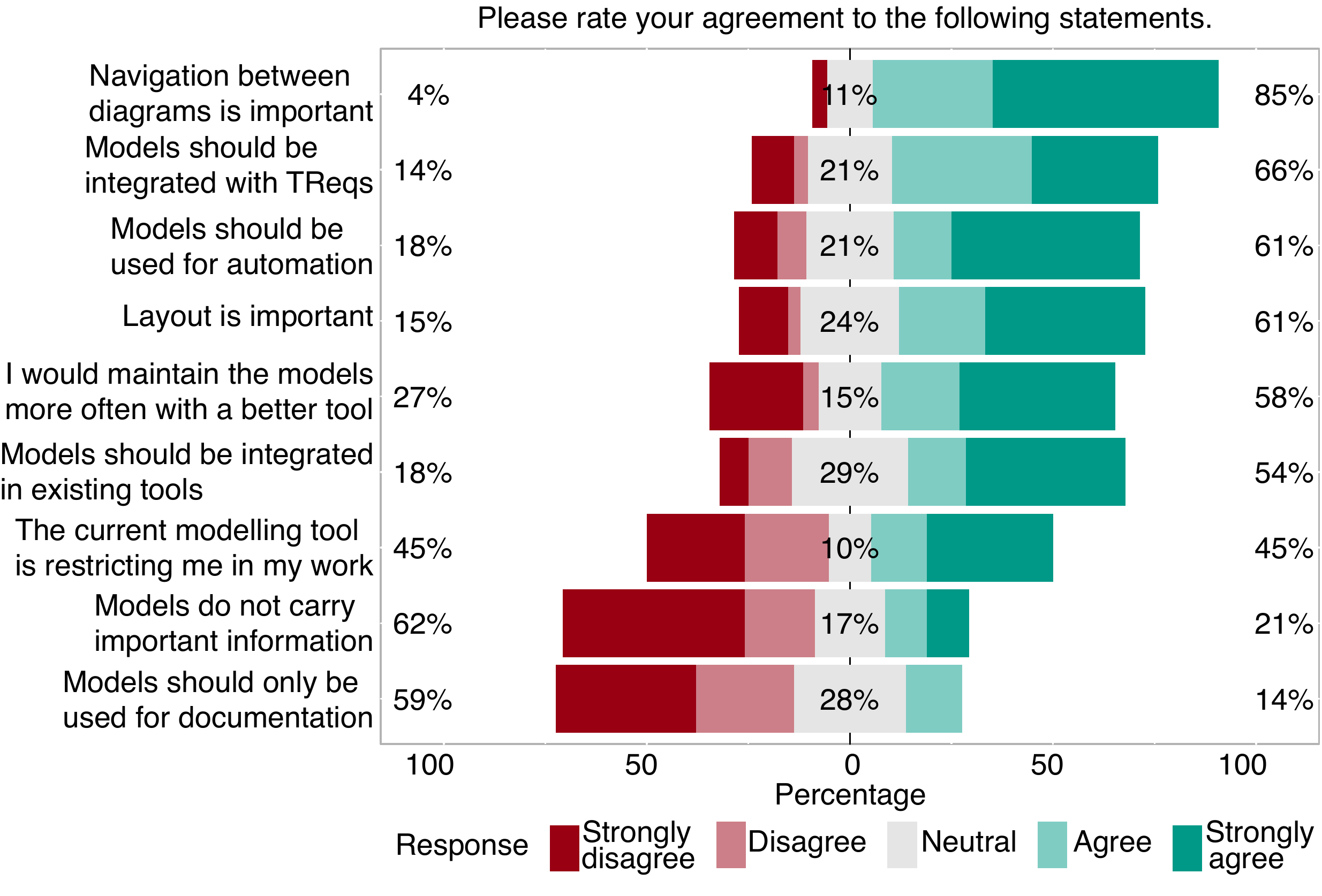}
    \caption{Agreement to Different Statements}
    \label{fig:agreementStatements}
\end{figure}

As expected from the study preparations, due to the domain, behavioural models dominate at the case company (Figure~\ref{fig:diagramUsage}). Activity (18 answers) and sequence diagrams (13 answers) are used by the majority of the participants. However, state machine and use case diagrams follow closely with 9 and 10 participants. Class and component diagrams are used by 4 and 3 participants only.

\begin{figure}
    \centering
    \includegraphics[width=.7\linewidth]{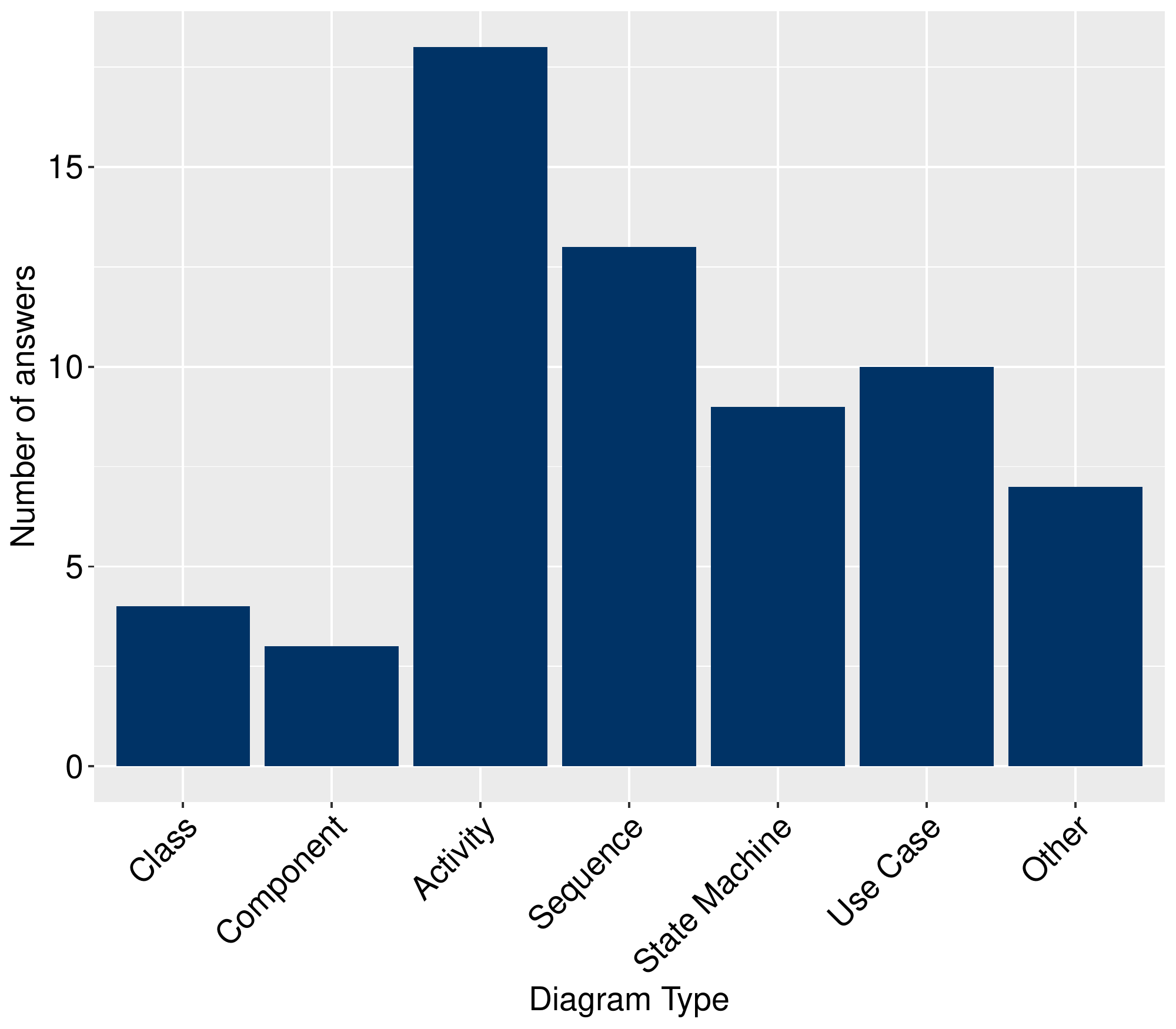}
    \caption{Diagram Usage According to UML Diagram Types}
    \label{fig:diagramUsage}
\end{figure}

For P6 (developers and testers do not think that the present models carry important information), it is rather interesting to observe that our proposition must be rejected based on our data (parts of Figure~\ref{fig:agreementStatements}). 
Indeed, 5 testers/developers out of 8 disagree or strongly disagree with the statement that existing models do not carry important information. 
Of the remaining 3 participants, only one agrees, with the other two being neutral or ``don't know''.

P7, the importance of layout, is clearly confirmed by our participants (parts of Figure~\ref{fig:agreementStatements}). 20 participants agree or strongly agree, 8 are neutral, and 5 (strongly) disagree.
One participant disagreeing noted that requirements should be stated in text, and have as a maximum pictures/models to support its explanation. Therefore, it should overall be kept simple, explaining their answer that layouting is indeed not important.

Regarding the integration of models into existing tools (P8), the picture is favourable (parts of Figure~\ref{fig:agreementStatements}). 
15 people (strongly) agree that this should happen, 8 are neutral, 5 against, and 5 don't know. 
Specifically, 19 participants agree that models should be integrated into the existing text-based requirements tool T-Reqs~\cite{knauss2018t}, with 4 disagree, 4 don't know, and 6 neutral answers.

From earlier work, we expected that informal modelling without any automation would be favoured by most stakeholders (P9). 
Interestingly, our results show a different mindset (parts of Figure~\ref{fig:agreementStatements}): 17 people agree that they should be used for automation, while 6 are neutral, and 5 each disagree or answered don't know.

Regarding P10, there is a disagreement as to whether the current solution is restricting the participants in their work (parts of Figure~\ref{fig:agreementStatements}): 13 each agree and disagree. 
4 don't know and 3 are neutral. While we initially thought that some participants could state that they are not restricted since they don't use the models, this picture was not confirmed clearly by looking at the disagreeing group: Only 2 of the participants stating that they don't write/modify models are in that latter group.

P11 again contradicted our impression from previous work - we expected that participants would state that they would not update their models even if the tool was better (parts of Figure~\ref{fig:agreementStatements}). 
However, 15 participants stated that they indeed would update the model if the tool was better. 
6 participants stated that they don't know, 5 were neutral and the remaining 7 disagreed.

P12, that navigation between diagrams is an important feature, got the strongest support in our survey (parts of Figure~\ref{fig:agreementStatements}). 
Indeed, 23 participants (strongly) agreed with the statement, 6 participants didn't know, 3 were neutral, and 1 disagreed.

\subsection{Survey Discussion}
Overall, we summarise the survey findings as follows.
We find strong support both for working with models (RQ1) and the use of requirements (RQ2) among our participants.
With respect to the needs to support the intended use of requirements models (RQ3), layouting and navigation, a focus on activity and sequence diagrams, and close integration into development tools and version control systems surfaced.
We discuss these aspects in this section and revisit them in the second, interview-based part of this study with the aim to shed more light on these aspects.

For RQ1 (Sentiments for and against models) we expected a diverse result, based on own experience and literature that propose a divide between model proponents and opponents. 
Instead, we find that most participants clearly see the value of models. 
Interestingly, there were a few voices mentioning that text-based requirements would be enough and that models are too complicated to handle. 
In particular, participants mentioned that at the case department, there is only little requirement work per team, which could easily be handled in text.

Similarly, the results show that the picture for RQ2 (How do different stakeholders use requirements models?) is far from the negative one we expected. 
While it is true that models are created by few people, and also mainly accessed by them, the majority of our participants sees the value of models and also the information contained in existing models. 
This covers all roles, including testers and developers that do not have an SM role in parallel. 
Furthermore, the testing and developer roles are far from negative towards modelling. 
The model creators/maintainers have substantial experience, though we do not know their educational background in modelling. 
Indeed, the move from the existing modelling tool to an integrated solution is supported, with few exceptions. 
From free-text comments, we see that there are several factors hindering the use of models at the moment. 
These include lack of tool access and tool usability, the complicated nature of models, the amount of details and need for constant work related to models, and the outdated information in models.
Finally, several statements related to process issues: Participants stated that there was currently no clear direction on whether the models should be kept updated, no process of doing so, and a lack of knowledge how to model and on which abstraction level. This means that the role of reading and understanding the model and then feed the information into the teams ends up in the hands of a few people (SMs). Participants suggested regular modelling courses for users, clear abstraction levels on what should/should not be in the models, and examples of models that are considered to be of high quality.

Regarding needs to support modelling (RQ3), participants strongly supported the notion that layouting and navigation are key features. Models at the case department often contain multiple requirements in flow charts/activity diagrams, with one requirement per activity node and a text description of each. The entire diagram then gives the context of the requirement, i.e., what happens before and after, and how it relates to other requirements. Often, there are links to other diagrams as well. Therefore, both the layout and the navigation are required to understand how the system behaves as a whole.
Our proposition was confirmed that only few diagram types are in active use, mainly activity and sequence diagram. However, there were minor usages of several further types.
Free text answers clearly pointed to the fact that any modelling tool needs to be integrated into daily work (e.g., into git), by using the same tools developers use and by being able to integrate the models with (text-based) version control such as git. While pictures are helpful, the models should theoretically be readable in text, in particular changes to models. Finally, a large share of the participants stated ease of use as one or the main success factor for a modelling tool.

\section{Confirmatory Interviews}
\label{sec:interviewProps}
After the survey, several gaps in our understanding remained, in addition to new questions that arose.
These gaps directly follow from the survey findings in relation to the propositions in \ref{tb:props} and \ref{tb:prop_eval}.
Since there is a general willingness to work with requirements and models, combined with a sense that current support is lacking, and clear indication of specific needs, clear questions for follow-up in-depth interviews follow (see Appendix~\ref{app:interviewProps}).

In the following, we discuss the findings relating to our three RQs.
Given the open nature of interviews, themes in the data can relate to more than one research question. 
First, we discuss how interviewees see the role of requirements in VLS agile systems engineering in Section~\ref{sec:reqsInVLS}, the role of models in VLS agile systems engineering in Section~\ref{sec:modelsInVLS}, and the use cases arising therefrom in Section~\ref{sec:use_cases}.
All these topics relate to RQ1 and RQ2.
Finally, we discuss the consequences for tooling (RQ3) in Section~\ref{sec:use_cases}.

\subsection{RQ1/RQ2: Requirements in VLS Agile Systems Engineering}
\label{sec:reqsInVLS}

\begin{tcolorbox}[float, title=Requirements and VLS Agile (RQ1/2)]
\begin{itemize}
\item System-level requirements are available too late in the process.
\item Requirements are an asset when changes are made, but often need to be updated first.
\item Importance of requirements: interviewees agree, but have doubts about general sentiment in organization.
\end{itemize}

Conclusion: Role of requirements in VLS agile is conceptually unclear.

\end{tcolorbox}

At the case department, requirements used to be written prior to development. 
Now, due to the agile transformation at the company, only vague requirements are developed prior to the sprints, which are then shaped and refined in parallel to the development and testing. 
Some interviewees perceive this as documentation work only, while others see it as a crucial step in invention and in preparing for future maintainability.
That is, the role of requirements in large-scale agile development is perceived very differently in the case department.
Several interviewees take the standpoint that there are too few requirements, and that those are written too late in the process.
They take the traditional development point of view in which upfront requirements analysis guides the development later on, and in which requirements provide the system knowledge.
The lack of such requirements is therefore seen as an issue.
\begin{quote}
{\em ``And we have them always too late in the chain. That's my view of it.''} -- Interviewee 1
\end{quote}

\begin{quote}
{\em ``[..] someone updates the implementation and suddenly things don’t work anymore. And then the problem is you have to determine why. Because a lot of the behavior of the product is not really based on requirements. We don’t have requirements on exactly everything.''} -- Interviewee 4
\end{quote}

None of our interviewees stated that they considered requirements unimportant. However, several of them did express that this was a common belief within the company. That is, that other sources but written requirements are sufficient to obtain system knowledge, e.g., test cases, or annotations to standards (compliance declarations).

\begin{quote}
{\em ``I got the feeling that some people think it's very important and some think this...we don't need requirements at all. We can do the coding and then we check at the end if it works OK, if the customer doesn't complain it's ok.''} -- Interviewee 1
\end{quote}

We therefore conclude that the notion of requirements in VLS agile systems engineering is conceptually unclear and individual opinions of practitioners differ.

\subsection{RQ1/RQ2: Requirements Models in VLS Agile Systems Engineering}
\label{sec:modelsInVLS}

\begin{tcolorbox}[float, title=Requirements Models and VLS Agile (RQ1/2)]
\begin{itemize}
\item Requirements models are important to understand the big picture.
\item Requirements models are hard to keep up to date.
\item Some models are increasingly outdated, thus losing value.
\item Changes can break a model and require re-design.
\item Different modelling styles make shared modelling difficult.
\end{itemize}
Conclusion: While requirements models provide substantial value, using them successfully in practice is challenging.
\end{tcolorbox}

Given that the role of requirements is conceptually unclear or at least different from the original, plan-driven process in which requirements were written up front, the role of using models to convey requirements information is also debated at the case company.

Several of our interviewees valued the existing requirements models.
They reported that the models serve primarily \emph{as a boundary object} between different \emph{agile islands} and the overall system, providing the long-term knowledge \cite{Kasauli2020}.
A common issue in VLS agile systems engineering is that individual methodological islands exist in a company that are disconnected \cite{kasauli17,kasauli21}, e.g., individual Scrum teams and an overall plan-driven process.
Having a model that relates system-level requirements to each other can help building bridges between the islands and keep knowledge over a long time.
For example, the models can help engineers understand how isolated user stories connect to the overall system behavior.
Furthermore, incoming change requests can be understood better in relation to the current system-level behaviour.
\begin{quote}
{\em ``The system model really like defines...[..] we put requirements that tell how something should behave in relation to some other functionality.''} -- Interviewee 2
\end{quote}

\begin{quote}
{\em ``Yeah. I think it's...if we [..] say that we have a graphical representation with flow charts and requirements and so on. I think that's very good.''} -- Interviewee 1
\end{quote}

\begin{quote}
{\em ``So it is, if you check the requirement there..so it doesn't give any relationship with other requirements. So you don't get...You just read it as a text, and you don't know actually how to relate to that. So the model, it is a complement.''} -- Interviewee 3
\end{quote}

However, we also have several interviewees that reject the requirements models, for several reasons.
First, while they consider requirements models useful in principle, they differ whether it is worth spending the required effort to create and maintain the models over time.
Just as with other forms of documentation, maintenance is essential.
If the model becomes outdated, it loses its value to the engineers.

\begin{quote}
{\em ``The problem is that the information gets outdated and there are not enough resources to make sure they are correct. And the focus probably lies on other things.''} -- Interviewee 1
\end{quote}

\begin{quote}
{\em ``We have definitely the knowledge to do the model right. The question is if we want to spend the time and effort. Because it would require many people many months to go through them all and update it.''} -- Interviewee 4
\end{quote}

In fact, an interviewee stated that several models at the company are outdated at the moment, and would require a substantial effort to be updated.

\begin{quote}
{\em ``The requirement model itself has degraded in many cases to the point where it's useless, totally inaccurate and not up to date.''} -- Interviewee 4
\end{quote}

Also in relation to maintenance effort, one interviewee stated that changes to the requirements can be orthogonal to the way the requirements models are designed, thus leading to substantial maintenance effort up to entire re-designs of a model.
For instance, changes that lead to a modification in the system structure could require moving requirements between models or entirely re-designing the information flow in models that depict behaviour or interactions.
\begin{quote}
{\em ``Because the model can be modelled in the way that it's hard to fit in my new requirement [..] And then I...okay, should I add to the mess [..] or should I try to remodel this? I think most people just add something with least effort and then the model becomes even harder to add something to.''} -- Interviewee 2
\end{quote}
While only one interviewee mentioned this issue, we considered it critical enough to list it here.

In addition to maintenance and model creation effort, several interviewees highlighted that there is no common way of modelling.
Currently, engineers do not get any instructions on how to create a model, how to use it, and how to maintain it.
This leads to a multitude of different modelling styles, reluctance to modify a model, and in many cases to teams abandoning the model altogether.
\begin{quote}{\em
    ``[..] there’s so much freedom. And we happen to have different styles depending on which person is doing the work. You try to...you like to model it in a way you like.''} -- Interviewee 5
\end{quote}

We further hypothesise that this also leads to a higher overall effort, since a person used to one model might need additional training to use or modify another model, as it might be modelled following a completely different style.

Finally, several interviewees have reservations towards requirements models due to tooling issues, and integration of the tools into the process.
Most of these reservations are similar to tool challenges known from related work, e.g., \cite{hutchinson11a,hutchinson11b,whittle13,liebel18sosym,liebel18survey}.
For instance, the interviewees mention outdated, heavy-weight tools, and the risk of vendor lock-in.

\begin{quote}{\em
    ``I have logged into Rhapsody just a few times, but in general that's very slow and so on. So that's not an option to log into it to get information.''} -- Interviewee 1
\end{quote}

\begin{quote}{\em
    ``Someone checks out the requirement document and only that person is allowed to make changes until it checked in. And hopefully that person will check it in before leaving for vacation.''} -- Interviewee 4
\end{quote}

Based on these themes, we conclude that requirements models provide substantial value in VLS agile systems engineering.
However, practitioners struggle to use them successfully due to challenges in maintaining them and modelling in a consistent style that allows engineers to work on shared models.

\subsection{RQ1/RQ2: Use Cases for Requirements Modelling}
\label{sec:use_cases}

\begin{tcolorbox}[float, title=Use Cases for Requirements Modelling in VLS agile (RQ1/2)]
\begin{itemize}
\item Models provide an overview of the requirements and their relationships.
\item Models provide valuable information to developers and testers.
\begin{itemize}
\item Many read, few write
\item Potential imbalance (effort/benefit)
\item Potential lack of awareness and appreciation of models
\end{itemize}\end{itemize}
Conclusion: A lightweight approach to requirements models that exposes models to many stakeholders is seen most favourable by the interviewees.
\end{tcolorbox}

As a third theme in our analysis, we discuss the different use cases for requirements models that our interviewees report or discuss, and the roles that relate to these use cases.
These are either already in place at the case company today, or the interviewees raised them as desirable or promising.
Not all interviewees had a good overview of all stakeholders that actually interact with the models, yet implicit assumptions on which roles should interact with the models existed.

The requirements models at the case department are primarily a collection of flow charts/activity diagrams.
Activities are used as containers for textual requirements and their connections depict the connections/traces between requirements.
There is typically a main flow and potentially multiple alternative flows, describing error cases.
The main value of the model lies in the overview it provides, primarily obtained through the relationships between requirements.
Several interviewees state that this overview is something that is hard to achieve with a text-only representation.
\begin{quote}
{\em ``I think that's very good. Because it's easy to follow, compared to when it's text based.''} -- Interviewee 1
\end{quote}

\begin{quote}
{\em ``It's impossible to read all of them and understand what the total requirement mass is.''} -- Interviewee 2
\end{quote}

The primary use case for the existing requirements models at the studied department is read-only access, to provide valuable information to developers to inform their activities.
However, in many cases this information is provided by other roles.
There is the widespread idea that mainly the \emph{SM} reads the model in order to then inform other roles and to provide an overview. 
SMs use the model as a source of information to answer questions regarding the overall system functionality, to investigate how changes affect the system, and to understand if change requests are due to misunderstood requirements, bugs, or actual changed needs.
This restricted use of the requirements models has the advantage that other team members do not need to be experts in modelling.
However, the disadvantage is that they might not be aware of the models' value and purpose, leading them to believe that updating the model is a waste of time - they do not see that the SM uses the model as a core element in their work.

The degree to which different SMs use the requirements models depends on their personal preferences and the state of the model.
As discussed earlier, some models are outdated and therefore no longer used by the SMs.

When discussing future use cases, most interviewees mentioned that all team members should read the model, but not necessarily write.
\begin{quote}{\em
    ``Everybody should at least have read access. I cannot see any reason why you should not have read access.''} -- Interviewee 1
\end{quote}
 
Relatively few stakeholders currently modify the model.
These are primarily the SMs, who create and update requirements models according to changes in the system, e.g., newly-implemented user stories.

Testers currently benefit from the requirements models to understand which requirements relate to a given object under test.
Again, the degree to which they use the models varies.
Interviewees also expressed that the model should allow testers better linking of test-related information, such as individual test executions.
This is currently not possible, but would enable better integration of work the testers currently have to do in other tools.

\begin{quote}{\em
    ``And what we as a tester see is lacking, is the way you work with requirements and the actual test executions, and how you follow up that your requirements are fulfilled.''} -- Interviewee 1
\end{quote}

\subsection{RQ3: Tool Features and Information Content in Requirements Modelling}
\label{sec:toolFeatures}

\begin{tcolorbox}[float, title={Needs for Requirements Modelling in VLS Agile (RQ3)}]
\begin{itemize}
\item Models need to be navigable and searchable.
\item Broad and easy access to the models is key to adoption.
\item Education and guidance for modelling need to be provided.
\item Review mechanisms similar to code reviews can foster adoption.
\item Generation artefacts from models can serve as a maintenance incentive.
\item Heterogeneity of teams and tools needs to be supported.
\item Automation of model layout is a trade-off.
\end{itemize}
\end{tcolorbox}

Based on the evidence from our interviews, several hypothetical ways to use requirements models exist.
We extrapolate the features necessary in tooling for requirements models, and the information content those models need to have.
Note that this section is only partially based on evidence from our interviews, and partially a logical extrapolation based on our expertise in the field.
For each theme, we discuss to what extent we do have evidence for the discussion points.

We distinguish four hypothetical scenarios based on our data. These are:
\begin{enumerate}
    \item Entirely abandoning requirements modelling in favour of using other artefacts as sources of information.
    \item Using requirements models as sources of information for the SMs only.
    \item Using requirements models as sources of information for developers, with SMs maintaining the model.
    \item Use and maintenance of the models by the entire development organisation.
\end{enumerate}

Scenario 1 (no requirements models) makes modelling tools unnecessary.
Hence, the tools do not need to be discussed.
\paragraph{Documenting Knowledge:}
Instead, abandoning models raises the question where the information should reside instead at the case department.
That is, information on how the overall requirements relate to each other, e.g., in terms of main and alternative flows/scenarios.
In our survey and during the interviews, we found several statements that existing documentation such as the user manuals could serve this purpose.

\begin{quote}{\em ``Yes, you could use [customer documentation] as a requirement if it works, but it has not always worked.''} -- Interviewee 5
\end{quote}

Similarly, tests are often raised as a potential source of knowledge that could replace written requirements, both in our data and in related work.
\begin{quote}{\em     ``Yes, actually I think it is an interesting idea because we have spent over the years quite a lot of time and effort on doing this requirements modelling. And there are alternatives which are tempting. Some have proposed that we should use [..] test cases as such, so instead we spend more time on reviewing the test cases and whatever ever changes we do to test cases, to see that this is still the wanted behavior.''} -- Interviewee 4
\end{quote}

However, our interviewees also raise concerns that tests might not be sufficient.
That is, each test expresses exactly one scenario, which means that the overall system behaviour arises from the combination of the entire test suite.
Therefore, this overall behaviour is not easily visible.
\begin{quote}{\em     ``If you only have the test case, it’s not clear really what parts that the test case verifies that our requirements can...and what is just a behavior. That’s a risk.''} -- Interviewee 4
\end{quote}

If requirements models are used in some capacity, several important needs arise.
Some of these are already present in the current tool solution at the case department, others are lacking according to the interviewees.

\paragraph{Supporting Traceability:}
In addition to the information being present, requirements need to exist so that testers know what to test, and have a target they can trace to.
Currently, the tool T-Reqs~\cite{knauss2018t} fulfils this purpose at the case department, even though one interviewee expressed that the possibilities for tracing are limited.
For instance, test executions could not be traced in T-Reqs and could therefore not be addressed in the tracing.

\begin{quote}{\em     ``I would like to say that this test execution, I will map to that requirement for our work package. To indicate that we have delivered what we are supposed to do and we are fulfilling this requirement. And then, two weeks later, the test execution fails. But [..] it means that somebody else maybe has destroyed, or we have delivered something new.''} -- Interviewee 1
\end{quote}

Hence, while traceability capabilities exist in T-Reqs, improvements are necessary.

\paragraph{Navigable and Searchable Information:}
In the context of VLS systems engineering, requirements and their relations quickly become complex.
Hence, it is important that they can be navigated and searched efficiently.

\begin{quote}{\em     ``We often have flows. So we have a number of requirements that are a part of a flow. So when something happens we follow a flow. But those flows are often broken down into sub-flows and the sub-flows might be re-used from other flows and things like that. So you want to have some way to kind of link it all together and make it easy to navigate. [..] you should be able to easily to search there and navigate.''} -- Interviewee 4
\end{quote}

\begin{quote}{\em     ``there is the practical thing of it. And for our requirements to be useful it has to be first of all easy to find and navigate. Because it’s a complex model, you can’t just...read one model, one short requirement out of context. [..] So you need an easy way to navigate the model.''} -- Interviewee 4
\end{quote}
For instance, links between requirements could be made navigatable by using hypermedia with hyperlinks between requirements, as is standard in most RE tools.
Currently, this is supported by T-Reqs for textual artefacts.
For models, several modelling tools allow for hierarchical models that support hiding information in sub-models, or distributing models and diagrams over several files.
However, extracting relevant information from models is difficult \cite{liebel18sosym}, e.g., in the form of a search.

\paragraph{Broad Information Access:}
Several of our interviewees stated that access to requirements information needs to be open to everyone.
If only selected roles have access to the information, requirements easily become an abstract concept that many engineers are not aware of or do not consider important.
This lowers the overall acceptance of requirements as an important source of knowledge in the organisation.
\begin{quote}{\em     ``Everybody should at least have read access. I cannot see any reason why you should not have read access.''} -- Interviewee 1
\end{quote}

\begin{quote}{\em     ``I think it’s important to be used and to be...maybe have good qualities, I think it’s good if it can be [..] easily accessed. That seems like a crucial point I think.''} -- Interviewee 2
\end{quote}

While easy tool access can help acceptance of the models in any case, especially for Scenario 3 and 4 (developers use the models at least as a source of information) this feature is crucial.
Previously, the case department used IBM Rhapsody, which required engineers to set up a remote environment to open the tool.
This turned out to be a large obstacle and only few engineers accessed the model, effectively limiting the access to the SMs.

\begin{quote}
{\em ``R: Because like two thirds of the people will not even try or \ldots Rhapsody. 
I don’t have that [remote] environment setup and the Rhapsody tool. I have never seen that tool [laughter].''} -- Interviewee 2
\end{quote}

\begin{quote}{\em     ``Well, first of all it has to be accessible, both for the people who need to do updates of the model, and also for the people then who are supposed to read the model, read the requirements, the designers and testers. I mean if it’s very easy to access, people will do it. If it’s hard to access, people will not.''} -- Interviewee 4
\end{quote}

\paragraph{Need for Education and Guidance:}

Similar to coding guidelines that exist in most organisations, \emph{modelling} and \emph{requirements} guidelines need to be in place that ensure a common approach to modelling and requirements.
While important for any kind of scenario in which requirements and models play a role, this guidance becomes more important when many people are supposed to edit models, i.e., for Scenario 4 in particular.
Several of our interviewees raise this point.
\begin{quote}{\em
    ``[..] everybody has access. But that of course means that there should be guidance. So when should you use it? And why should you use it? And who should use it for what reason?''} -- Interviewee 1
\end{quote}

\begin{quote}{\em
    ``[..] people don’t spend effort on making it correct. So you cannot really trust it, because people maybe not have the ambition or maybe not the knowledge [..].''} -- Interviewee 2
\end{quote}

\begin{quote}{\em
    ``Yeah, it is back again to this if we want to use Rhapsody and the modelling, in that sense that we call it modelling, then we have to use the modelling guidelines. And keep it. Not just know it and just abuse the model.''} -- Interviewee 3
\end{quote}

As a variant of guidelines, several interviewees also suggest mentors at the company that can support others in modelling-related questions.

\begin{quote}{\em
    ``If they don’t know it, double-check with someone that knows it.''} -- Interviewee 3
\end{quote}
\begin{quote}{\em     ``So need to be one or a couple of people that really know how to model that you can ask for 'Okay, can we have an hour and come to a conclusion how we should model this, my problem?'.''} -- Interviewee 2
\end{quote}

\paragraph{Need for Review Mechanisms:}
While guidance and education can improve the quality of models, enforcing standards could become difficult.
Hence, mechanisms are required to do so.
Drawing from experience both in RE and in programming, we believe that both reviews (similar to code reviews) and automated analyses (similar to requirements heuristics or static code analyses) have the potential to enforce model quality.
Reviews are also brought up several times by interviewees as a reason to rely on textual modelling tools like PlantUML.

\begin{quote}{\em     ``[..] have like reviews in the tool or it would be like very much be a good way to get people into also use it, the others. Of course then you can easily ask them to 'Okay, can we have a review of this?' or 'Can you review this?'.''} -- Interviewee 2
\end{quote}

\begin{quote}{\em     ``But if you instead did it in a pure text, then you can use your standard merge tools. You can do standard diff to see what has been updated. If someone wants to do a review, I mean we have today code review tools we use for everything else.''} -- Interviewee 4
\end{quote}

\paragraph{Code Generation: Carrot and Stick:}
Existing models at the case company are in many cases outdated or of low quality.
While guidelines, mentors, and model reviews could help addressing this, several interviewees suggested that artefacts could be generated from the models in a form of lightweight model-driven engineering process.
This would encourage people to read and update the models, as they serve as a ground truth.
In turn, generation could help enforcing guidelines.
For instance, using elements that are semantically wrong would lead to incorrect artefact generation or errors during the generation process.

\begin{quote}{\em
    ``[..] maybe it can just generate a text. [..] I know that in Rational Rose you could generate a Word document. ''} -- Interviewee 3
\end{quote}

\begin{quote}{\em     ``Either they have to have a demand. They have to check it, otherwise they cannot do the job. Or you give them, I don’t know, a carrot. [..] Because modelling is taking time. If you want to do it right. [..] if the modelling tools could give you something that you could generate, then it would be a little bit better, I suppose.''} -- Interviewee 3
\end{quote}

Due to the use of the requirements model as a read-only source of information, the model is disconnected from the final product.
Nothing is generated from the model that is used further in development.
This means that the end product could potentially be completely contradicting the model.
This bears the risk that the model deteriorates.
If code or other important artefacts would directly be generated from it, this would not happen.

However, it is important to note that generation of artefacts requires a well-balanced approach:
None of the interviewees expressed the desire to follow a strict generation approach, where, e.g., the entire code is generated based on models.
Hence, generation should remain a tool that allows engineers to see some direct benefit of the models, and to obtain quick feedback of sorts, without dictating their entire workflow.

\paragraph{Supporting Heterogeneity:}
In VLS systems engineering, development work is organised in a number of different ways.
For instance, component teams are common, where different teams focus on their individual components.
The case department instead structures work by expertise, e.g., having teams that focus on quality attributes such as availability.
This heterogeneity leads to different approaches, and to different needs.
Furthermore, heterogeneity and independence of teams is encouraged by the use of agile practices.
In turn, this heterogeneity requires highly flexible tools and notations, or independence within the teams to choose their own tools and notations for requirements and requirements modelling.
Forcing a single tool/style/approach on all teams will likely lead to resistance.
Nevertheless, advertising success stories from individual teams might pave the way for others to adopt similar approaches.

\begin{quote}{\em ``We have different areas with different needs, but my area is very functional.''} -- Interviewee 4
\end{quote}

With respect to requirements modelling, supporting heterogeneity might also mean accepting that some teams choose to abandon modelling entirely, either due to preference, or due to a mismatch with their way of working.

\paragraph{Abstraction Level:}
The case company needs to be compliant with specifications from 3GPP \cite{3gpp_tr1601}.
This standard contains many technical details, that often need to be discussed or referenced in the requirements, e.g., to discuss required additions.

\begin{quote}{\em     ``[..] somewhere I would say between 75\% and maybe up to 90\% depending on a bit what area we are in, the requirements are specified by 3GPP, we implement the standards. And then we don’t need to re-specify this. But sometimes we need to clarify this because the standard might not be very clear on certain details. So we might need to annotate it and say 'okay, in this case the value should be this or we do like that', when the standard is not clear enough.''} -- Interviewee 4
\end{quote}
This leads to requirements or requirements models on a low level of abstraction.
Any notation or specification format used at the case company needs to support this level of abstraction.
For example, to make a requirements model understandable, it might be necessary to use hierarchical models to hide details.
A modelling tool would thus need to offer sophisticated hierarchy and decomposition features.

However, given that the requirements at the case department are closely-aligned with the standard, this also means that engineers have an easier time choosing the right level of abstraction, something that is otherwise challenging~\cite{liebel18sosym}.

\paragraph{Importance of Layout:}
Finally, several of the interviewees raise the advantage of textual modelling languages like PlantUML, as they can be integrated into traditional text-based environments such as git or diff.
However, they come at the cost of relying on automated layouting, e.g., through GraphViz\footnote{https://graphviz.org/} in the case of PlantUML.
The layout of a graphical model can contain important information, and reflect the intent of the modeller \cite{storrle2011impact}.
Expressing this information is therefore no longer possible with automated layouting.
Our interviewees have differing views as to whether this is a problem or not.

\begin{quote}{\em     ``I don’t think that will be any problem. [..] It could be better like that. I see it, as long as I can show the flow, you have space to write, you could show the relation of the different parts, I think it should be okay.''} -- Interviewee 5
\end{quote}

\begin{quote}{\em     ``I think that's crap [automatic layout]. You must be able to...an automatic thing is good, but then you should click on a button 'I want an automatic suggestion'. And then you should be able to fine-tune it and it should stay that way. [..] Because the tool will never know what my intention was.''} -- Interviewee 1
\end{quote}

One interviewee suggested that, while automatic layout might not work for complex models, conventions or adjustments to automatic layout could be possible.

\begin{quote}{\em     ``[..] it’s not a black and white question here. In many cases it doesn’t matter. If you do modeling as diagrams or activity flows or something like that...if you have small enough flows the layout doesn’t matter, because it will not be that bad. [..] When the model becomes larger and more complex layout becomes more important. [..] it might be good to do some hinting. For example exceptional flows maybe should go to the right while main flows on the left or something like that. ''} -- Interviewee 4
\end{quote}

This last point clearly shows the complex trade-off between the simplicity of tools, and features that might be considered essential by some. 
\section{Discussion and Conclusion}
\label{sec:discussion}
We conducted a case study in one department at Ericsson AB, a large Swedish telecommunications provider, investigating the use of models for RE purposes.
We conducted a survey with 33 participants, followed by 5 semi-structured interviews.

With respect to RQ1 (What sentiments exist for and against the use of requirements models in VLS agile systems engineering?), we find that our study participants consider requirements models useful and valuable.
While several interviewees mentioned that sentiments against these models exist in the case department, we did not directly interact with anyone supporting this view.
However, we also find that creating and maintaining requirements models at a sufficiently high level of quality is challenging.
Several participants maintain that many existing models have deteriorated over time and are no longer useful.
Additional point worth highlighting is that different modelling styles make it difficult to jointly work on models, something that might be harder to unify compared to, e.g., writing style in textual requirements.
Finally, one interviewee mentioned that the nature of some changes to requirements models can require entire models to be re-drawn.
This either leads to substantial overhead, or it causes resistance to make changes in the models, especially if changes are made by other engineers than the model creator.

The use of existing requirements models at the case department (RQ2, How do different stakeholders use requirements models in VLS agile systems engineering?) is primarily by SMs, in their role as providers of system-level knowledge and as a boundary between incoming change requests, system requirements, and work in individual agile teams.
While several engineers in other roles use the models as well, mainly in a read-only fashion to answer their questions on intended system requirements, complex tooling used in the case department in the past has prevented a broader consumption of the models.
Interviewees expressed the desire that all engineers should at least read the models.
They further express confidence that their tool T-Reqs supports this, in which models are stored in textual format alongside code and textual requirements in git repositories.
This allows easy access through tools engineers use on a daily basis, as well as easier review in terms of textual diff.

Our findings with respect to RQ1 and RQ2 allow for reasoning about the information content and tooling needs (RQ3, What are the needs to support the intended use of requirements models in VLS agile systems engineering?) regarding requirements modelling.
The value of the models at the case department is primarily in providing an overview of the system requirements and especially their connections, something that is difficult to express in a suitable manner in text.
However, in order for the models to reach their full potential, the contained information needs to be up to date and of high quality.
This, in turn, requires a broad access to the models by all stakeholders, at least in read-only fashion.
Furthermore, education and guidance in how to create and maintain the models is essential, potentially also in the form of mentors at the case department.
In terms of tool features, navigation and search are essential.
Furthermore, interviewees expressed the desire to incorporate the requirements models in their regular code review workflow, e.g., by adding them to the git repository in textual form.
Using generation of artefacts from models could be a way to further incentivise their use.
However, no interviewee expressed the desire to generate substantial code from the models.
Using textual models is considered an advantage by most study participants, due to easier tooling and the possibility to integrate the models into exiting tools.
However, a number of issues arise due to the reliance on textual modelling, most notably the loss of manual layout capabilities at the case department.
Interviewees suggested workarounds such as a standard layout, where the main flow is always displayed to the left, while alternative flows are drawn to the left of the main flow.
Finally, we note the importance of supporting heterogeneity at the case department, with different needs and preferences in the agile teams.

Our study clearly shows the usefulness of models during RE, if used for well-motivated use cases.
Furthermore, the study shows that simple modelling tools that are close to the engineers in terms of workflow and tooling have the potential to be successful, while heavy-weight modelling tools do not reach their full potential due to difficulties in accessing and using the tools, and resistance to do so regularly.
Finally, we find several trade-offs that exist when tailoring models and modelling tools to an organisation, e.g., sophisticated modelling tool features such as hierarchy and manual layout vs. simple, text-based modelling tools. 

\bibliographystyle{ACM-Reference-Format}


\section{Acknowledgments}
We would like to thank our contacts at Ericsson AB for the fruitful collaboration and constructive input at all stages of this work, and the study participants for their valuable contributions. Furthermore, we express our gratitude to João Araujo for feedback on the manuscript draft.
Parts of this work were supported by Software Center Project 27 on RE for Large-Scale Agile System Development.
\appendix

\section{Questionnaire}
This section shows the questionnaire we ran with the case department.
\subsection{Page 1}
\textbf{Rationale:}\\
In this questionnaire, we aim to investigate the current use of requirements models in [Case Department], as well as the needs for future use. The reason to do so is the planned replacement of IBM Rational Rhapsody. At the same time, this questionnaire is part of an academic research project at Chalmers \& Gothenburg University in collaboration with [Case Company], investigating the use of models for requirements engineering in industry.\\

\noindent \textbf{Target group:}\\
We target all individuals at [Case Department] that come into contact with requirements models. This means both people who create models and those who access them.\\

\noindent \textbf{Practicalities:}\\
The survey consists of 20 questions and will take approximately 15 minutes to answer. Your answers are treated completely anonymous. The survey ends on [end date].\\

\noindent Thank you very much for your time!\\

\noindent If you have any questions, please contact the creators of this survey: Grischa Liebel (grischa@chalmers.se) or Eric Knauss (eric.knauss@cse.gu.se) for any research-related questions, and [Contact Person 1] or [Contact Person 2] for any [Case Company]-related questions.

\subsection{Page 2}
\begin{enumerate}
\item What is your main role? \\ \textbf{(Free text)}
\item In which system area do you work? \\ \textbf{(Free text)}
\item How long have you been working with [Product]? \\ \textbf{(Free text)}
\item What are your main work tasks? \\ \textbf{(Options, multi-selection)}
\begin{itemize}
\item Line management
\item Requirements specification
\item Architecture definition
\item Design definition
\item Software implementation
\item Testing
\item Customer support
\item Security
\item Quality management
\item Process improvement
\item Organisation improvement
\item Other:
\end{itemize}
\end{enumerate}
\textbf{Models}\\
In the following, we refer to "models" in the sense of UML or similar notations.

\begin{enumerate}
\setcounter{enumi}{4}
\item How much experience do you have in creating models (in years)? \\ \textbf{(Mandatory, free text)}
\item How often do you on average create/modify models of requirements? \\ \textbf{(Mandatory, single-selection: Never, Yearly, Monthly, Weekly, Daily)}
\item How often do you on average read/use models of requirements without modifying them? \\
\textbf{(Mandatory, single-selection: Never, Yearly, Monthly, Weekly, Daily)}
\item How often do you on average access (modify/read) models of requirements created outside your team? \\
\textbf{(Mandatory, single-selection: Never, Yearly, Monthly, Weekly, Daily)}
\item Do you personally use IBM Rational Rhapsody for creating/reading models of requirements?  \\ \textbf{(Mandatory, Yes/No)}
\item Do you have any additional comments on this page?  \\ \textbf{(Optional, free text)}
\end{enumerate}

\subsection{Page 3}
\begin{enumerate}
\setcounter{enumi}{10}
\item Which of the following statements correspond to your understanding of the terms diagram/model? \\ \textbf{(Mandatory, Multi-selection)}
\begin{itemize}
\item There is no difference between a diagram and a model.
\item A model contains all entities and relations, while a diagram is a (partial) \item visualisation of this model.
\item Multiple diagrams can exist for the same model.
\item A diagram always has a graphical representation, while a model may not.
\item Other: \textbf{(Free text)}
\end{itemize}
\item Are you in general positive or negative towards the use of models to express requirements? (Note that this question is not used to rate your answer. Instead we aim to find out what the general view of models is in the organisation.) \\ \textbf{(Mandatory, 
single-selection: Positive, Negative, Neutral/I don't know)}

\item Which diagram types do you use (creation or usage) for expressing requirements? \textbf{(Optional, Multi-selection, with an example diagram for each diagram type)}
\begin{itemize}
\item Class diagrams
\item Component diagrams
\item Activity diagrams
\item Sequence diagrams
\item State Machine diagrams
\item Use Case diagrams
\item Others (Free text)
\end{itemize}

\item Please rate your agreement to the following statements. \\ \textbf{(Mandatory, 5-point Likert scale from Strongly Disagree to Strongly Agree, plus an "I don't know" option)}
\begin{itemize}
\item The layout of a diagram is important to me.
\item The existing requirements models do not carry important information.
\item The access to requirements models should be integrated with existing development tools.
\item The access to requirements models should be integrated with T-Reqs. Requirements models should be used for automated tasks (e.g., verification, artefact generation).
\item Requirements models should only be used for documentation.
\item The current modelling tool is restricting me in my work.
\item I would update/maintain the requirements models more frequently if I had a better tool.
\item Navigation between diagrams is an important feature.
\end{itemize}
\item Do you have any additional comments on this page?  \\ \textbf{(Optional, free text)}
\end{enumerate}

Additionally, we displayed the following two questions iff a participant stated in question 6 (model creation) that he/she created/used models at least monthly:
\begin{enumerate}
\setcounter{enumi}{15}
\item For what purposes do you create models of requirements?  \\ \textbf{(Optional, free text)}
\item Who is looking at the models of requirements you create?  \\ \textbf{(Optional, Multi-selection)}
\begin{itemize}
\item Myself
\item Other developers in the same team
\item Other teams
\item Product owner for my team
\item Product owners of other teams
\item CPI writers
\item System managers
\item Customer support
\item Others:
\end{itemize}
\end{enumerate}

\subsection{Page 4}
\begin{enumerate}
\setcounter{enumi}{17}
\item What features/properties does a modelling tool need to offer?   \\ \textbf{(Optional, free text)}
\item Currently, keeping the models/diagrams of requirements up to date is challenging. Do you have any suggestions how engineers could be motivated to maintain these models/diagrams more frequently?  \\ \textbf{(Optional, free text)}
\item How would the ideal situation regarding requirements modelling look like in the future?  \\ \textbf{(Optional, free text)}
\item Do you have any other comments (e.g., alternative ideas to modelling, tool suggestion)?  \\ \textbf{(Optional, free text)}
\end{enumerate}

Additionally, we displayed the following question iff a participant stated "yes" in question 9 (use of IBM Rhapsody):
\begin{enumerate}
\setcounter{enumi}{21}
\item Which features in IBM Rational Rhapsody are important to you?  \\ \textbf{(Optional, free text)}
\end{enumerate}

\subsection{Page 5}
\textbf{Thank you for completing this questionnaire!}\\

\noindent We would like to thank you very much for helping us.\\

\noindent Your answers were transmitted, you may close the browser window or tab now. 
\section{Proposition Evaluation}
\label{app:prop_eval}

\begin{table}[!ht]
\centering
    \begin{tabular}{|l|l|}
    \hline
    Prop Nr & Used Questions \\
        \hline
        P1 & Q1, Q4, Q6, Q7, Q8, Q9, Q17\\
        P2 & Q1, Q4, Q6, Q7, Q8, Q9, Q17\\
        P3 & Q5, Q6 + free text\\
        P4 & Q1, Q4, Q6, Q7, Q8, Q12, Q14.2.7\\
        P5 & Q13 + free text\\
        P6 & Q1, Q4, Q14.2\\
        P7 & Q14.1\\
        P8 & Q14.3.4\\
        P9 & Q14.5\\
        P10 & Q14.6.7, Q12, Q9\\
        P11 & Q14.7\\
        P12 & Q14.8\\
                \hline
    \end{tabular}
    \caption{Evaluation of Propositions}
    \label{tb:prop_to_q}
\end{table}

\section{Interview Propositions and Questions}
\label{app:interviewProps}
In the following, we discuss the updated propositions and open questions guiding the interview design.
The open questions relate in particular to contradictions in the survey data.
Propositions are accompanied with \emph{How}-questions, as we aim to understand their outcome in detail, not simply corroborate them or not.

\subsection{Propositions}
\begin{itemize}
\item The current ad-hoc use of models is insufficient. Either modelling should be abandoned, or a clear process (with clear stakeholders, tasks and abstraction levels) and guidelines (including courses on modelling) are needed. (\emph{How could such a process look like?})
\item Information in models is outdated in many areas. This needs a centralised effort to be fixed, replacing the tool does only treat the symptoms. (\emph{How could the case company proceed? How do first steps look like?})
\item A number of stakeholders/tasks have been forgotten when considering the use of models and the tool integration. (\emph{Who are these stakeholders? What are their needs?})
\item Potential users need a clearer motivation for using (and in particular updating) models. (\emph{How could we motivate them?})
\item A lightweight modelling approach is sufficient for the case company. They require only very few model elements of activity diagrams (activity nodes with text, relations between them) and few model capabilities. (\emph{Which of the features of modern modelling tools are still required?})
\end{itemize}

\subsection{Questions}
\begin{itemize}
\item Using PlantUML allows only automated layouting. How do stakeholders view this trade-off between text-based integration into T-Reqs, and losing capability to modify the layout? How does the simple approach relate to other modelling capabilities?
\item Automation using existing models was supported in the survey. How should this automation look like? What aspects could/should be automated?
\item What kind of information is needed in the models? What is a suitable level of abstraction?
\item Working with models is currently cumbersome. How can the experience be improved? What does it mean to be easy to use for a modelling tool?
\item Information needs clearly differ between stakeholders. What information needs exist for specific stakeholders?
\end{itemize}

\section{A-Priori Codebook}
\label{app:codebook}
\begin{itemize}
\item role, experience (demographic codes)
\item modellingProcess, stakeholdersOfModels, informationContent, motivationForModels, trade-off, abstraction, modelAutomation, toolFeatures
\end{itemize}
\end{document}